\documentclass{nature}
\usepackage{graphicx}
\usepackage{amssymb}
\usepackage{braket}

\title{Quantum-Inspired Computing: Can it be a Microscopic Computing Model of the Brain?}

\author{Yasunao Katayama \\
IBM Research - Tokyo \\
{\tt email: yasunaok@jp.ibm.com \\}}

\begin{document}

\maketitle
\begin{abstract}
Abstract: Quantum computing and the workings of the brain have many aspects in common and have been 
attracting increasing attention in academia and industry. The computation in both is parallel 
and non-discrete. Though the underlying physical dynamics (e.g., equation of motion) 
may be deterministic, the observed or interpreted outcomes are often probabilistic. 
Consequently, 
various investigations have been undertaken to understand and reproduce the brain on the basis 
of quantum physics and computing \cite{BQCO, BQC, BQC2, BQC3, Holo, BQC4, BQC5, BQC6, BQC7}. 
However, there have been arguments 
%from physics and cognitive science points of view 
on whether the brain 
can and have to take advantage of quantum phenomena that need to 
survive in the macroscopic space-time region at room temperature
\cite{antiBQC, antiBQC2, antiBQC3}. This paper presents a unique microscopic computational model 
for the brain based on an ansatz that the brain computes in a manner similar to quantum computing, 
%not with quantum waves 
but with classical waves. 
Log-scale encoding of information \cite{KataUCNC} in the context of  
computing with waves \cite{Kata_Tran} is shown to play a critical role in bridging the computing models with  
classical and quantum waves. 
Our quantum-inspired computing model opens up a possibility of unifying the computing framework of 
artificial intelligence and quantum computing beyond quantum machine learning approaches.
\end{abstract}

\newpage

\tableofcontents

%\singlespacing

\newpage

\section{Introduction}

Artificial intelligence (AI) and quantum computing (QC) are two rapidly evolving technologies that 
are redefining computing. The capability of computers to handle narrowly defined AI tasks is 
surpassing human capability, and the research focus is shifting 
toward giving computers broader and more general AI capabilities \cite{AIS}. 
Quantum computers are expected to solve certain types of problems that conventional 
computers are hard to solve \cite{QC}. 
%AI systems are evolving and surpassing human capabilities for domain-specific tasks. 
%Still, in terms of efficiency metrics such as energy per computation, 
%there exist significant gaps between the computer and the brain. 
%Since Quantum Computing (QC) and the workings of the brain have many aspects in common, 
%their relationship has been attracting increasing attention in academia and industry. 
Since QC and the workings of the brain share many aspects in common, 
it is not surprising to see that 
there have been various research activities aimed at understanding the computing model 
of the brain on the basis of quantum physics and 
QC as early as the 1970s \cite{BQC, BQC2, BQC3, Holo, BQC4}. 
As a result of rapid advancement of QC, quantum machine learning 
has been attracting a lot of attention in both AI and QC research communities \cite{BQC5, BQC6, BQC7}
Indeed, the computation in both cases is executed in parallel with non-discrete variables. 
Though the underlying physical dynamics (e.g., equation of motion) 
may be deterministic, the observed or interpreted outcomes are more probabilistic. 
%Consequently, various investigations have been undertaken to understand and mimic 
%the brain on the basis of quantum physics and computing. 

However, at the same time there have been controversial arguments from physics (i.e., material) and 
cognitive science (i.e., mental) points of view on whether the brain, both biological and artificial, 
can and have to take advantage of quantum features, such as Tensor-product statespace and 
entanglement, that need to survive in macroscopic 
space-time regions at room temperature \cite{antiBQC, antiBQC2, antiBQC3}. 
Though macroscopic quantum phenomena are increasingly being observed in the lab \cite{MacroQ, MacroQ2}, 
they are seldom perceived in ordinary life. In addition, 
from a computing point of view, though there is much less doubt on inherent QC advantages in 
quantum information processing \cite{Feynman, Bennett}, AI workloads in general process huge data sets and  
irreversible algorithms with noise in computing environment and data, 
which may not always be effectively translated into tensor product entangled statespaces with a 
limited number of qubits and their interconnects. 
In other words, the statespace advantage of QC may quickly 
diminish unless the symmetry of the problem cleanly fit 
to exponentially-large Hilbert state spaces with exclusively linear operators on them. 
QC-unique constraints such as no cloning, strict reversibility, 
measurement complications, may pose additional challenges in application coverage, 
system scalability, and fault and error tolerances, 
if we dare to pursue strict QC approaches universally for classical and data-intensive 
computing problems of AI workloads. 
On the other hand, parallel 
and non-discrete dynamics is not unique to quantum physics but is something natively 
observed in ordinary classical physics, such as wave dynamics \cite{Kata_Tran}. 
%The phase coherence time 
%of these classical waves is typically much longer ( $>$ 1 s) than the quantum counterpart 
%on the macroscopic scale at room temperature. 

%The work reported here has been motivated by the question of whether energy-efficient and highly 
%integrated brain-inspired 
%computing systems can be built using classical waves with orders of magnitude slower group velocity 
%$v_g \ll c$ \cite{Kata_Tran}, where $c$ is the speed of light. In other words, whether dynamics and computing 
%with spike trains 
%in the brain can be modeled as classical slow wave processes.  
%As shown in Fig. \ref{fig1}, the approach we take in the present paper is 
%to define a quantum-inspired physical computing model of the brain beyond the QC boundary, 
%which greatly differs from the conventional approaches in this area, i.e., 
%defining a brain computing model within QC. 
%% Our approach is also different from holographic approaches \cite{Holo}
%Our approach unique in starting with clarifying which QC features can be 
%exploited to model the workings of the brain, rather than fully accepting QC features. 
%%A computing model of the brain is developed that computes in a manner similar to quantum computing, 
%%however, not with quantum waves but with classical waves. 
%%The model is proposed as a base-line physical computing model of the brain that can unify 
%%AI and QC. 

\begin{figure}
\centering
\includegraphics[width=9cm, clip]{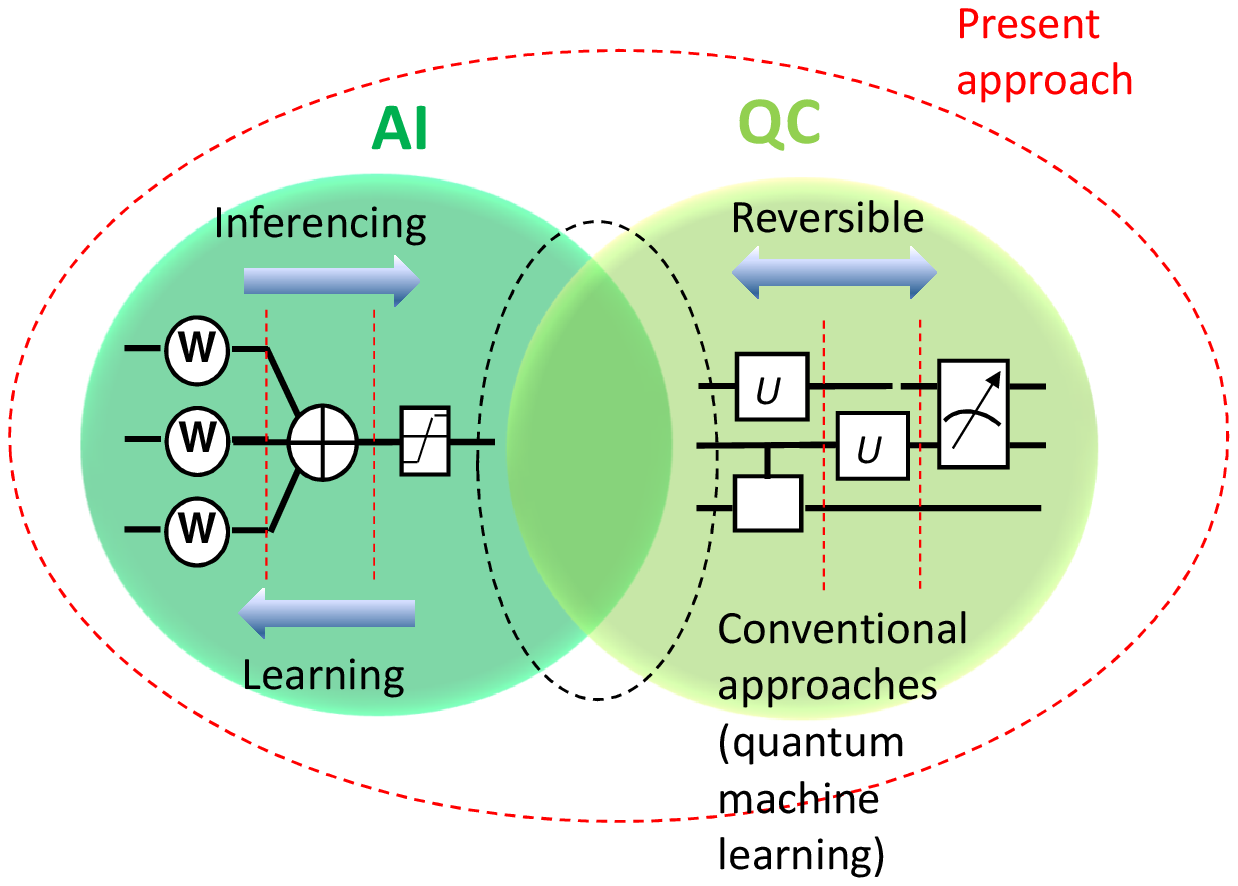}
\caption{Approaches for physical computing models of the brain. In the proposed approach, 
quantum-inspired approach for AI is 
investigated, ultimately aiming at unifying AI and QC computing models with classical and quantum waves.}
\label{fig1}
\end{figure}

Here in this paper, we propose quantum-inspired computing (QIC) as energy and 
computationally efficient microscopic computing model for classical 
workloads, in particular AI.   
As shown in Fig. \ref{fig1}, 
our approach is unique compared with existing QC-based approaches, 
%(including quantum-classical hybrids \cite{QCH}), 
since quantum physics is not prerequisite in the computing model. 
Instead, parallel computing advantage of QC is incorporated in a more general 
natural computing context with less QC-unique constraints. 
More specifically, our model is based in as a network of elastic wave processing of spikes. 
Wave superposition and thresholding can smoothly and precisely execute, 
without being influenced by collision and congestion, weighted sum and ReLU operations, 
which are key arithmetic in modern AI processing. 
Our quantum-inspired modeling approach with natively incorporating precise spike correlations 
in continuous time allows for the temporal 
computing more rigorously based on original spiking neural network (SNN) model \cite{Maass}
and should differentiate it 
from existing neuromorphic approaches \cite{NC, NC2, NC3}. 
In order for physically directive and nonblocking   
wave nature to prevail at a small scale for energy efficiency and integration, 
it is essential to reduce the wavelength by reducing the spike signal velocity \cite{Kata_Tran}. 
Quantum-inspired approach has been improving classical counterpart at different levels \cite{Rec}. 
The present work is quantum-inspired approach at a microscopic computing level. 

\section{Encoding cubits into classical waves}
The wave nature of matters and fields plays a critical role in both classical and quantum physics. 
In classical physics, a wave is a phenomenon that can elastically transfer energy and information, 
without transporting matter, via the interplay between displacement and a linearly responsive 
restoring force, such as between a displacement current and an electrical and magnetic field in the 
case of electromagnetic waves. Waves carry electrical signals, not electrons. 
In quantum physics, a wave function is an essential 
means to describe quantum phenomena in both ground and 
excited states. 
Though classical waves do not exhibit quantum-mechanical coherence, 
such as macroscopically observed in lasers and superconductors, 
they have their own wave characteristics such as superposition and interference. 
Classically coherent waves can be generated by externally exciting the system in phase 
with forced vibrations. Such classical waves are in accordance with classical dynamics 
with commutable variables and contiguous energy spectra. 
Interestingly, the coherence of such macroscopic classical waves can often last for 
a substantial amount of time even at room temperature, as observed in radio waves in 
wireless communication or sound waves in a music hall. 

\begin{figure}
\centering
\includegraphics[width=9cm, clip]{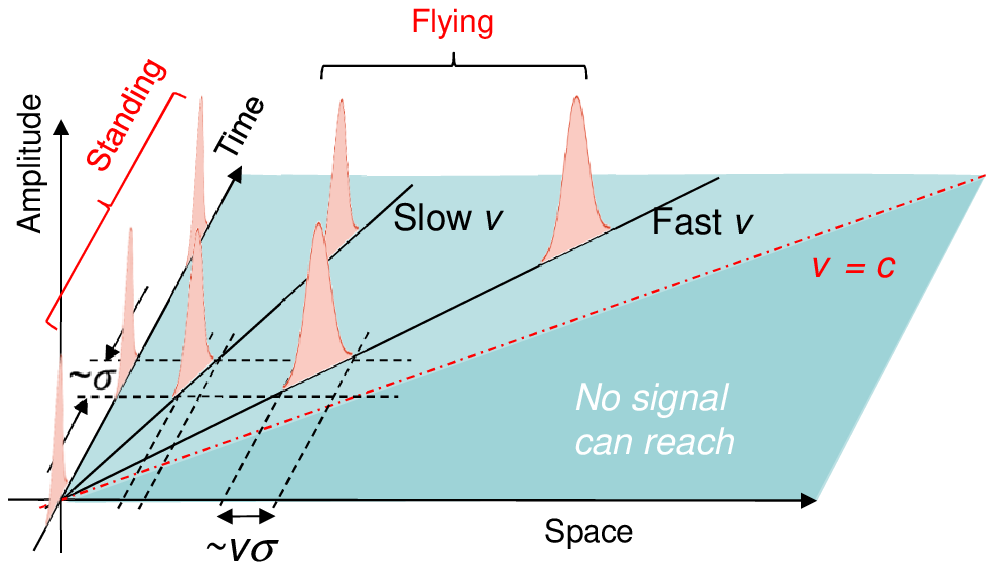}
\caption{Flying vs. standing cubits. Theoretically the distinction may be  
superficial since it is just a matter of the coordinate selection. 
Slower $v$ results in 
smaller spatial area $v\sigma$ for each spike wave packet with 
the temporal width of $\sigma$ and thus better integration and energy 
efficiency \cite{Kata_Tran}. The line $v=c$ represents the light cone in the 
Einstein-Minkowski space time. The original spatial width at standing is ignored for simplicity.}
\label{fig2} 
\end{figure}

In this paper, we will construct a computing model with classical waves.  
Let us start the discussion on how to take advantage of the classical wave features 
in computing by defining, in analogy to the qubit in quantum computing, 
the cubit, which is an abbreviation of {\it classical universal bit}. 
The notation used for cubits is similar to the standard Dirac notation for qubits but with 
double bras and kets. Let $\ket{\ket{0}}$ and $\ket{\ket{1}}$ be normalized orthogonal 
basis vectors for cubits: 
\begin{equation}\braket{\braket{0|0}}=\braket{\braket{1|1}}=1,\end{equation}
\begin{equation}\braket{\braket{0|1}}=\braket{\braket{1|0}}=0.\end{equation}
The scalar product is defined as a conjugate integral in a predefined volume $V$, 
which can be flying or standing as discussed later. States with higher indices and harmonics 
for cubits can be considered similar to qubits \cite{HH}, but we will not go further 
for simpler comparison. Using 
\begin{equation}
\braket{\braket{x_1|x_2}}=\delta{(x_1-x_2)} 
\end{equation}
enables the actual waveform in space time to be given as
\begin{equation}f_0(x) = \braket{\braket{x|0}},\end{equation}
\begin{equation}f_1(x) = \braket{\braket{x|1}}.\end{equation}
Therefore, the scalar product can be expressed as an integral in space coordinates: 
\begin{equation}\braket{\braket{0|1}} = \int_{x \in V} f_0^*(x)f_1(x)dx.\end{equation} 
The difference between bra or ket may not matter much if the waves are defined in $\mathbb{R}$. 
Complex conjugate $*$ should be used when dealing with complex numbers. An arbitrary cubit state 
\begin{equation} 
\ket{\ket{a}} = \bar{a} \ket{\ket{0}} + a \ket{\ket{1}}
\end{equation}
can have a wave form in space time as 
\begin{equation}
a(x) = \bar{a}f_0+af_1.
\end{equation} 

Cubits can be flying or standing depends on whether the basis vector $\ket{\ket{0}}$ and $\ket{\ket{1}}$  
and the associated volume $V$ is flying or standing. 
The difference in the mathematical formulation is not as large as that in 
the implementation, given that it is only a matter of which coordinate to 
select, as illustrated in Fig. \ref{fig2}. 
In other words, when a flying cubit and associated $V$ are moving at a constant velocity $v$ ideally with 
little dispersion, the flying cubit 
can be considered standing when viewed from the coordinate attached to it. 
Assuming $v \ll c$ (i.e., nonrelativistic) 
with no dispersion, an arbitrary flying 
%qubit or 
cubit state 
%$\ket{a(x,t)}_f$ or 
$\ket{\ket{a(x,t)}}_f$ can be related to a standing one 
%$\ket{a(x.t)}_s$ or 
$\ket{\ket{a(x,t)}}_s$ as 
%\begin{equation}
%\ket{a(x,t)}_f = \int \ket{a(x-x',t)}_s  \delta(x'-vt)dx',
%\end{equation} 
%or
\begin{equation}
\ket{\ket{a(x,t)}}_f = \int \ket{\ket{a(x-x',t)}}_s \delta(x'-vt)dx'.
\label{eqq}
\end{equation} 
The use of slow $v$ is essential to exploit wave nature with energy efficiency and integration 
\cite{Kata_Tran}. 
In QC systems, 
qubits are often standing. Flying qubits are diffusive except for massless qubits, i.e., those 
flying at the speed of light. 
On the other hand, spikes in the brain are more consistent with flying cubit picture. 
Encoding information into standing cubits may be related to digital recording techniques, 
such as partial response maximum likelihood (PRML) encoding \cite{PRML}. 
While those techniques are considered as taking 
advantage of standing wave nature of recording signals with the finite frequency response 
of the media {\it for memory and storage}, 
the brain is considered as using flying wave nature of spike signals 
{\it for computing} in the present model.

There are multiple types of logical cubits:
$$
\begin{array}{lll}
{\bf Normalized ~ full ~ cubit} & \ket{\ket{a}}:= \bar{a} \ket{\ket{0}}+a \ket{\ket{1}}, 
|\bar{a}|^2+|a|^2=1 
& \in U(1) {\rm ~or~} SO(2) \\
{\bf Normalized ~ half ~ cubit} & \ket{\ket{a}}:= a \ket{\ket{1}}, 0 \leq |a|^2 \leq 1 
& \in U(1) \cap \mathbb{R} \\
{\bf Unnormalized ~ full ~ cubit} & \ket{\ket{a}}:= \bar{a} \ket{\ket{0}}+a \ket{\ket{1}} 
& \in \mathbb{R}^2 {\rm ~or~} \mathbb{C} \\
{\bf Unnormalized ~ half ~ cubit}. & \ket{\ket{a}}:= a \ket{\ket{1}} & \in \mathbb{R} 
\end{array}
$$
Cubits are defined not in the Bloch sphere ($SU(2)$ or $SO(3)$) like qubits. 
The $\ket{\ket{0}}$ amplitude for the half cubits is implicit and may result in 
more efficient gate implementation if $\ket{\ket{0}}$ can be regarded as the ground state 
(i.e., no excitation). 

\begin{figure}
\centering
\includegraphics[width=8cm, clip]{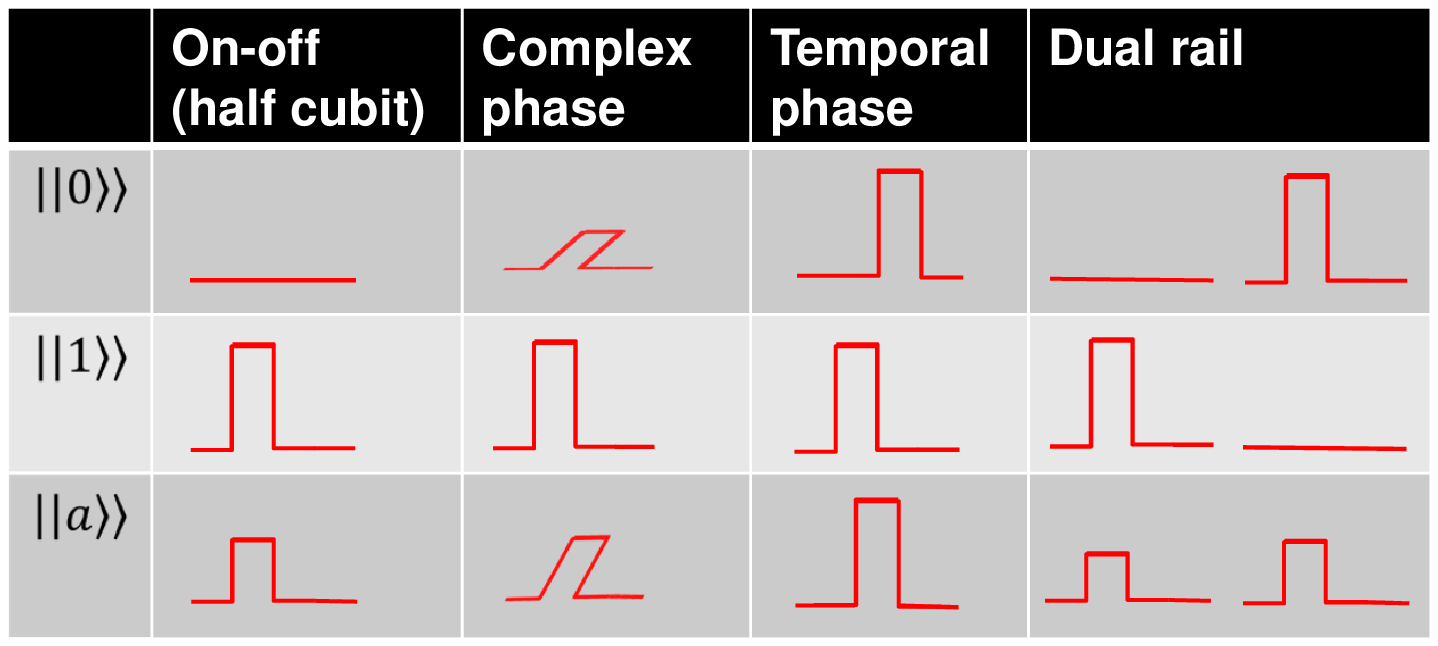}
\caption{Physical cubit encoding examples: (a) on-off encoding, (b) phase encoding, 
(c) temporal encoding, (d) dual-rail encoding. To be more consistent with biological spike signaling, 
RZ rather than NRZ is assumed.}
\label{fig3}
\end{figure}

There are several ways to encode logical cubits into physical waves. 
Specific examples are shown in Figs. \ref{fig3} (a)--(d). 
Examples (a) and (c) are for half cubits, and examples (b) and (d) are for full cubits. 
Examples (a)--(c) are with single wires, while example 
(d) is with a spatially distinguishable wire pair. 
In other words, examples (a)--(c) require a single lane per cubit, while example 
(d) requires dual-rail encoding, i.e., two physical lanes to encode a single logical cubit. 
The unnormalized half qubit could be ill defined without dual rail coding, 
since the amplitude of $\ket{\ket{0}}$ cannot be 
estimated from that of $\ket{\ket{1}}$. 
Example (b) is identical to phase shift keying in wireless communication, 
which is typical with a carrier. Example (a) can be considered as the real part of example (b) 
though (a) can be encoded without the carrier as well. 
Logical information can be encoded into multiple physical cubits by using majority 
logic coding, such as rate or population coding, or by using temporal interval coding 
on a continuous non-discrete time scale. 

Note that the present cubit formulation is mostly for computing, not for physics. 
We inherited the standard bra and ket 
notation from quantum physics to describe cubits in a consistent manner with qubits. 
If we dare to 
approximately relate cubit states to boson qubit ensemble states by taking the classical limit of the 
occupation numbers for $\ket{0}$ and $\ket{1}$, i.e., $n_0, n_1 \to \infty$, 
and by neglecting the off-diagonal terms, we get
\begin{equation}\ket{\ket{0}} \sim \sqrt{\braket{n_0}},\end{equation} 
\begin{equation}\ket{\ket{1}} \sim \sqrt{\braket{n_1}}.\end{equation} 
Therefore, cubits may be regarded as population coded qubits in ensemble average. 
This relation is to be revisited later for both bosons and fermions.  
%Fermion qubits have to involve other degrees of freedom in order for the population coding not 
%to be washed out due to Pauli's exclusion principle. 

\section{Operations for cubits}
\begin{figure}
\centering
\includegraphics[width=12cm, clip]{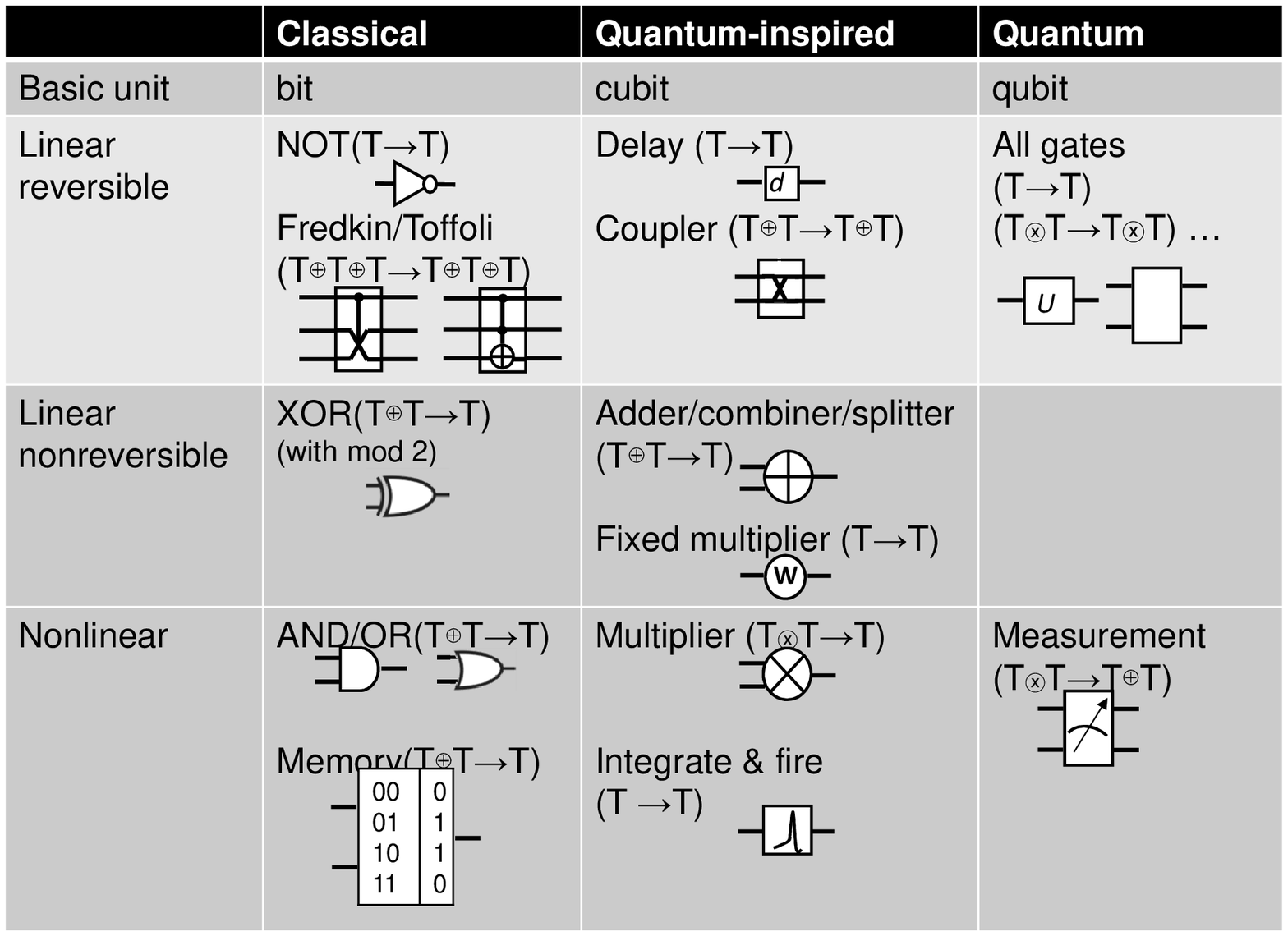}
\caption{Comparison of linear and nonlinear operations for single- and multiple-input 
primitives in classical, quantum-inspired, and quantum computing. $T$ represents mathematical 
space corresponding to the number system of choice for each basic unit.}
\label{fig4}
\end{figure}

QIC can have various gate primitives for linear and nonlinear operations. 
Linear operations are natural outcomes of wave superpositions, and the corresponding primitives are 
coupler with full cubits for reversible operations, 
and adder or combiner (and coupler with half cubits) 
for irreversible operations, respectively. 
Splitter is the reverse operation of combiner. 
The fixed multiplier with the constant $W$ 
for unnormalized cubits has to be either lossy ($|W|<1$) or active ($|W|>1$). 
Nonlinear operations such as wave multiplication are also possible. 
These operations for classical waves may look familiar to those with wireless 
communication backgrounds since they are typically used for splitting and combining, and for up- and 
down-conversion of wireless signals. Integrate and fire operation makes discrete binary decisions 
of either fire $\ket{\ket{1}}$ 
or not fire $\ket{\ket{0}}$ for normalized half qubit out of non-discrete continuous states $\ket{\ket{a}}$. 

Figure \ref{fig4} compares linear and nonlinear operations for single- and multiple-input 
gates with bits, cubits, and qubits in classical, wave-based, and quantum computing. 
The mathematical space corresponding to the number system of choice for each basic unit 
is represented by $T$. For example, $T=SU(2)$ for qubits. 
For classical computing, NOT and XOR gates perform linear operations over 
Cartesian product states 
$T \oplus T$ 
while AND and OR gates perform nonlinear operations. 
Fredkin or Toffoli gates are listed as classical reversible gates \cite{FT}.
The most general form of a nonlinear operation can be represented as a memory, 
for which any output can be defined at the expense of $2^N$ resources for $N$ address inputs. 
QC gates perform linear operations in SU(2) or in its tensor product space, 
$T \otimes T.$ 
The measurements nonlinearly reduce qubits in the tensor product states 
to ordinary bits in the Cartesian product 
states, though they are still projective linear operations in the original tensor product states, 
but nonlinearity can arise if multiple qubit measurement results are reduced in Cartesian product states. 
Measurements in QC can be related to integrated and fire in QIC as a decision process of superposed 
states to one of the orthogonal states.

\section{Log-scale encoding to bridge QC and QIC}

Let us consider a decision problem with each decision represented by 
a qubit: 
%or cubit: 
$p_i = |\braket{1|p_i}|^2$ 
%or $p_i = |\braket{\braket{p_i|1}}|^2$ 
and $\overline{p_i} = 1-p_i = |\braket{0|p_i}|^2$. In other words, 
\begin{equation}
\ket{{\tt In}_i} = \ket{p_i} = \bar{p_i}\ket{0}+p\ket{1}.
\end{equation}
%or $\overline{p_i} = 1-p_i = |\braket{\braket{p_i|0}}|^2$ 
When the state consists of a tensor product state of independent (i.e., no entanglement) qubit states as 
$\ket{p_0}\otimes\ket{p_1}\otimes ... \otimes \ket{p_n}$, 
the probability $P_{\{d_1 ... d_n\}}$ of having
a decision $\{d_1 ... d_n\}$ with $d_i = 0$ or $1$ is 
\begin{equation}
P_{\{d_1 ... d_n\}} = \prod_{j=1}^n |\braket{1|p_j}^{d_j} \braket{0|p_j}^{1-d_j}|^2.
\end{equation}
When we take the log of both sides, 
%The log-scale encoding translates a qubit relationship in a Tensor product into a cubit 
%relationship in a Cartesian product: 
\begin{equation}
\log P_{\{d_1 ... d_n\}} = 2 \sum_{j=1}^n \log |\braket{1|p_j}^{d_j}\braket{0|p_j}^{1-d_j}|.
\end{equation}

When $w_{ij}$ identical qubits 
%or cubits $j$, representing the number of replicated neurons and/or synaptic connections, 
are involved in decision $P_i$, the equations become
\begin{equation}
P_{i\{d_1 ... d_n\}} = \prod_{j=1}^n |\braket{1|p_j}^{d_j} \braket{0|p_j}^{1-d_j}|^{2w_{ij}}.
\end{equation}
and 
\begin{equation}
\log P_{i\{d_1 ... d_n\}} = 2 \sum_{j=1}^n w_{ij} \log |\braket{1|p_j}^{d_j} 
\braket{0|p_j}^{1-d_j}|.
\end{equation}
Note that no cloning theorem does not allow the duplication for arbitrary qubit states. 
 
The following log relationship can translate Tensor-product qubit to Cartesian-product cubit as:
\begin{equation}
\braket{\braket{1|{\tt Out}_i}} \sim \log P_{i\{d_1 ... d_n\}}, 
\end{equation}
\begin{equation}
\braket{\braket{d_i|{\tt In}_i}} \sim \log \braket{1|p_i}^{d_i} \braket{0|p_i}^{1-d_i}.
\end{equation}
In other words, the probability in a cubit are interpreted as 
a log scale encoding of the probability in a qubit.  
Log-scale encoding is a standard technique in probability calculations, such as 
used in log likelihood estimations. It simplifies probability calculations for a wide 
dynamic range inputs and seems consistent with that fact that the biological brain can compute with 
a wide dynamic range of sensory signals \cite{WF}. 
The encoding with excitatory and inhibitory neurons for $p$ and $\overline{p}$ 
in a dual-rail manner is possible. The log scaling encoding can also provide 
a natural rectifying capability with appropriate biasing $b_i$ as   
\begin{equation}
\braket{\braket{1|{\tt Out}_i}} = 2 \sum_{j=1}^n w_{ij} \braket{\braket{d_j|{\tt In}_j}} + b_i.
\label{eq}
\end{equation}
When the probabilities are normalized per unit time interval, they corresponds to the 
spike rate which is the inverse of the time-to-spike interval \cite{NS}. 
The neuron dynamics in this model is driven by the digital number of 
replicated spike paths rather than the analog strength of each spike and its synaptic weight. 
The log-scale encoding can significantly
reduce the number of spikes for a given signal dynamic range. 
%As a result of log encoding, 
%$\ket{\ket{Out_i(t)}} = \log P_{i\{d_1 ... d_n\}}$ and 
%$\ket{\ket{{\tt In}_i(t)}} = \log |\braket{\braket{p_j|1}}^{d_j} \braket{\braket{p_j|0}}^{1-d_j}|$.
When cubits are flying, Eq.\ref{eq} can be expressed, by using Eq. \ref{eqq}, as 
\begin{equation}
\ket{\ket{{\tt Out}_i(t)}} = \int \sum_j W_{ij}(t') \ket{\ket{{\tt In}_j(t-t')}} dt' + b_i,
\end{equation}
by substituting $W_{ij}(t) = 2 w_{ij} \delta(t-d_{ij})$, where $d_{ij}$ is the 
delay between node $i$ and $j$ and is related 
to the path length $L_{ij}$ by $d_{ij}=L_{ij}/v$.  
%This formula is consistent with the prior art \cite{Kata_Tran}.

\begin{figure}
\centering
\includegraphics[width=14cm, clip]{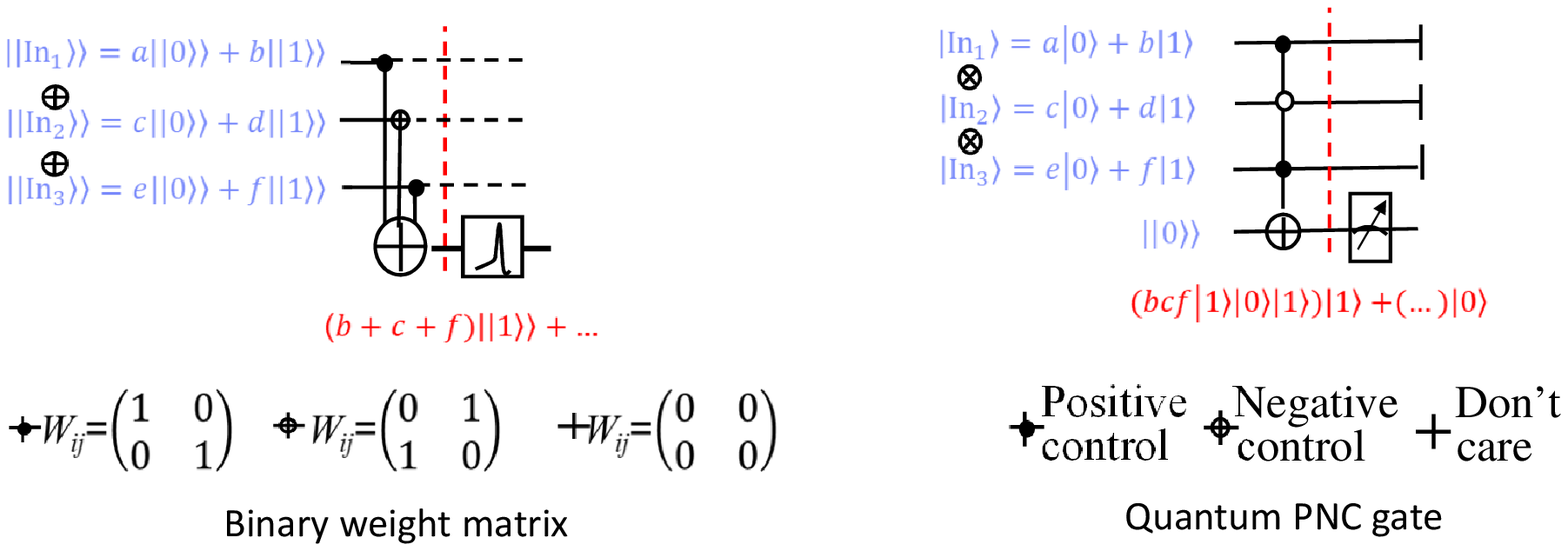}
\caption{Comparison of cubit and qubit operations for an artificial neuron with binary weight matrices 
in Cartesian and Tensor product 
states, respectively.}
\label{fig5}
\end{figure}

Comparison of cubit and qubit operations for a binary neuron as an example 
is shown in Fig. \ref{fig5}. Probabilities of cubits and qubits for a binary neuron can be inter-related 
by log encoding. 
The present model describes inferencing. 
%since learning is not a mandatory function for QC. 
Eventually, weight update rules for learning has to be worked out, 
ideally based on direct temporal correlations of spikes (i.e., cubits) 
rather than indirect formulations based on rate coding 
\cite{Hinton, Bengio}.

\begin{figure}
\centering
\includegraphics[width=12cm, clip]{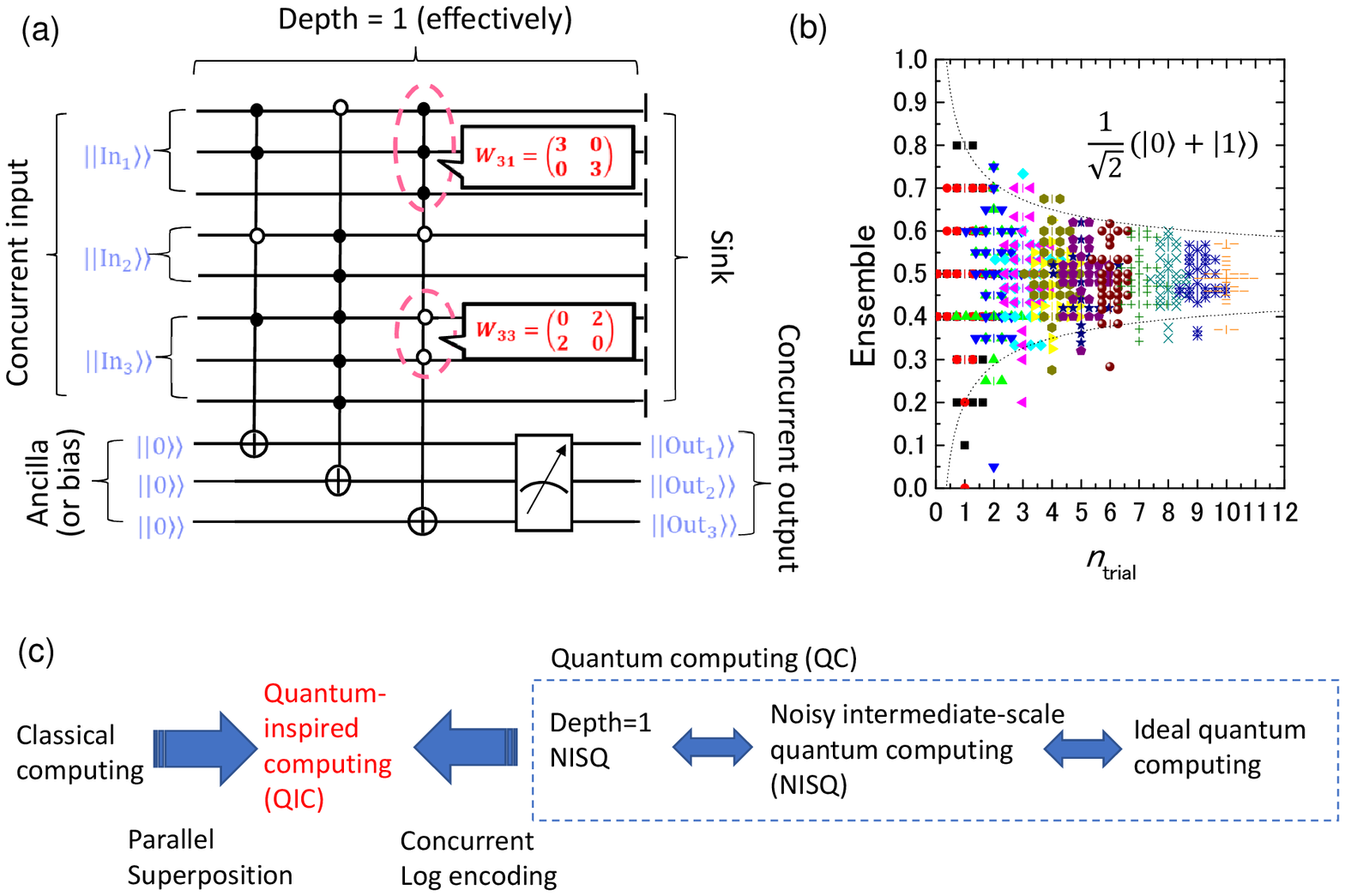}
\caption{(a) Neural operations in QIC interpreted as ensembles of parallel 
and concurrent QC operations; (b) Calculated result for ensembles as a function of $n_{trial}$; 
(c) QIC distinguish itself from classical computing in terms of 
parallel execution and superposed states and from QC in terms of log-scale encoding and 
concurrent execution.}
\label{fig6}
\end{figure}

QIC  
can become a superior one for non-quantum applications, so should not always be considered 
as an inferior version of QC. 
Indeed, the approach is aiming at expanding the application coverage of QC-like 
computing by appropriately loosened constraints of QC without much affecting its advantages. 
This quantum-inspired approach can be interpreted as depth=1 noisy intermediate-scale quantum 
computing (NISQ) \cite{NISQ} with ensembles of flying qubit 
inputs as spikes in density matrices. 
QIC for neural operations interpreted as ensembles of parallel 
and concurrent QC operations is shown in Fig. \ref{fig6} (a).
It can support nonbinary weights by the cubit inputs as 
ensembles of concurrent and parallel qubit inputs:
\begin{equation}
\ket{\ket{{\tt In}}}\bra{\bra{{\tt In}}} \sim n_{tral}(\bar{p}\ket{0}\bra{0}+p\ket{1}\bra{1}).
\end{equation} 
A calculated result for ensembles as a function of $n_{trial}$ is shown in Fig. \ref{fig6} (b). 
Each trial consists of ten parallel qubit measurements.
The variance is expected to converge as $\sim 1 / \sqrt{n_{trial}}$ for cubits by concurrently 
executing entire trials in parallel. This concurrent and parallel trial is valid for 
both bosons and fermions, and we can write as 
\begin{equation}
n_0 = n_{trial} \bar{p},
\end{equation}
\begin{equation}
n_1 = n_{trial} p.
\end{equation} 
As summarized in Fig. \ref{fig6} (c) QIC distinguish itself from classical computing in terms of 
parallel execution and superposed states and from QC in terms of log-scale encoding and 
concurrent execution. 
Since cubits are classical consisting of 
ensembles of orthogonal qubits, 
they can be duplicated to constitute nonbinary weights without being constrained by the no cloning theorem.  
Concurrent ensembles of try and measure executions can significantly accelerate and scale 
a wide range of less entangled AI workloads 
without QC-unique constraints even under noises in data and environment. 
Thus, the present QIC approach has a good potential 
to deliver much better throughput and scalability 
for the main-stream AI workloads than strict QC counterpart. 

\begin{figure}
\centering
\includegraphics[width=16cm, clip]{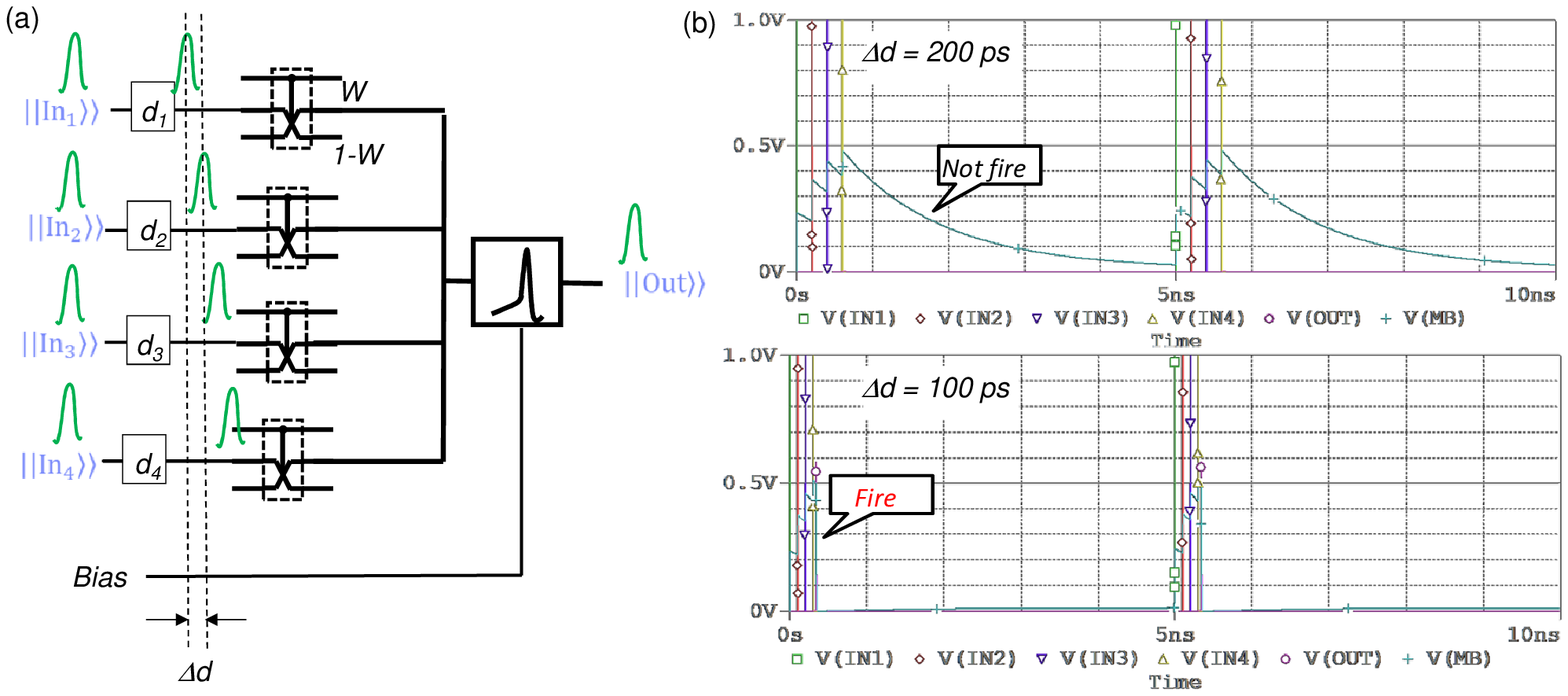}
\caption{(a) The diagram of an example artificial neuron 
based on the proposed computing model; (b) Simulation results with 
$\Delta d = $ 200 ps and 100 ps for upper an lower graphs, respectively. By appropriately adjusting 
the leakage time constant of the membrane potential MB, input spike temporal correlations can be extracted. }
\label{fig7}
\end{figure}

When these primitives are combined, neuron models to describe an abstract level PHY can be constructed,
as exemplified in Fig. \ref{fig7} (a).
It consists of wave processes in continuous time model.
The output activation is triggered by input correlations, not input-output (scattering model).
%Reciprocal learning path, in particular global one for supervised learning, 
%is a critical challenge in neuromorphic hardware.
It can accept both excitatory and inhibitory inputs. Weight matrix is expressed by Fredkin gate
as a coupler with weight adjustment,
details of which are described in Appendix A.  
Fredkin gate can output the leftover wave energy from the other port. This otherwise wasted wave energy 
would be useful for weight update with learning.  
%Full-fledged learning, not like reservoir. 
Simulation results is shown in Fig. \ref{fig7} (b) 
using a neural channel and circuit modeling technique discussed in another paper \cite{GLOBECOM, NANO2019}.  
When the leakage parameter of integrate and fire block is appropriately set, spike temporal 
correlations can be extracted by appropriately adjusting the leakage time constant of the membrane potential.  
When $\Delta d = d_{i+1}-d_i = $ 200 ps (upper), 
spikes from multiple inputs are less temporally correlated, and 
thus output does not fire. 
When $\Delta d = $ 100 ps (lower), spikes are more temporally correlated and output fires. 

\section{Discussion}
We have shown that computing with cubits can be performed 
in parallel by using superposed states, similar to qubits. 
Thus, here we address the differences between QC and QIC, 
starting with the advantages of QIC for classical 
computing tasks. 
In short, cubits are better fit for accelerating general classical algorithms. 
Using 
cubits can avoid the complications associated with reversibility, cloning, and measurement. 
This is important given that many real-world algorithms (e.g., those for AI tasks) are not reversible 
and require copying of data. in addition, there is less 
overhead for data IO when using cubits since information in classical domain does not 
have to be converted into information in quantum domain in Tensor-product states with a smaller number of qubits. 
The dephasing time for 
classical waves is often much longer even at room temperature, which is also the case for spike signals 
in the biological brain. 
The use of cubits thus offers the potential for deeper circuits with 
less error correction overhead. 
In contrast, QIC is not suitable for quantum information processing. No exponential 
tensor-product linear state spaces are natively available, and nonlocal entanglement 
features are missing. Even though both cubits and qubits can be reduced to the same 
classical bit operations, they behave differently in their own mathematical spaces. 
This makes it fundamentally difficult to use cubits to emulate qubits for quantum information processing. 
The following arguments in Appendix B may provide a better understanding of this situation 
by illustrating the extent to which cubits can be used to perform algorithms involving entanglement. 
More rigorous formulation of QIC may 
provide a new perspective on computational complexity arguments \cite{CC}.

Our approach is not as extreme as reversible computing \cite{RC} in terms of 
energy per computation limit, since it allows for and 
asks for dissipatively losing non-essential information. 
Thus, it can alleviate logical complications associated 
with stringent reversible computing constraints and more 
suitable for inherently irreversible AI workloads. 
We exploit elastic waves rather than matters 
(i.e., not billiard balls) as information carrier for stable solid-state implementation. 
The use of orders of magnitude slower waves than the speed of light is essential 
for energy efficiency and integration. Our conjecture is that this is what the 
biological brain has already taken advantage of with spike signals for energy efficiency.

Compared with existing neuromorphic computing approaches, our approach can 
better incorporate time as a resource for computation and communication, 
which is considered essential for SNNs \cite{Maass}. 
Otherwise, though spike signaling can improve tolerance to noise and disturbance 
including baseline wonder, simple rate coded spiking neural networks would result 
in exponentially less efficient than binary coded analog neural networks 
in terms of coding efficiency in neural channels \cite{GLOBECOM}. 
Thus, though many prior art neuromorphic research results claim energy efficiency, 
in reality the total power reduction would be limited when spiking IOs are inefficiently coded. 
We expect that, in addition to coding efficiency, supporting temporal coding, such as interpulse-interval 
coding \cite{IPI}, natively in continuous time with desired stochasticity without being affected by 
undesired temporal jitters due to signal collision and congestion can provide a new perspective. 
Importance of temporal delay in biological brains has also addressed in neuroscience \cite{PC}.

The proposed QIC should not be considered the same as standard analog signal processing.
The underlying physical processes are quite distinctive. 
The former is elastic and nondissipative while the latter is often inelastic and dissipative.   
In other words, the signal is represented by nonequlibrated wave dynamics, rather than equilibrated potential. 
It can be considered as a miniaturized version of computing with radio frequency (RF) passives, 
exploiting wave-based signal processing features, such as superposition, 
naturally without signal collision nor unwanted leakage. Signals are 
physically and logically directive and nonblocking to each other thanks to the wave nature.
As was mentioned earlier, it is essential to reduce the wavelength of the spike signal 
with respect to the transmission length by reducing the spike signal velocity.
This aspect has not been addressed much in conventional analog neuron models. 
%Further details 
%focusing on neuron channel model is to be presented else where \cite{GLOBECOM}. 

QIC can become a fair candidate 
for the microscopic computing model for the artificial brain, in particular considering to realize SNNs
with a good temporal degrees of freedom. It also provides a new perspective on 
the microscopic computing model for the biological brain whether quantum effects are mandatory to explain 
brain computing functionalities, in addition to quantum effects in biochemistry 
perspectives. From the computing model point of view, entanglement aspects would be  
a key to identify whether quantum effects are mandatory in order to explain brain functionalities. 
Further arguments on entanglement with QIC is given in Appendix B. 

\section{Conclusion}
We proposed quantum-inspired computing as energy- and 
computationally-efficient microscopic computing model for 
AI workloads in analogy to quantum computing. 
Parallel computing advantage with superposed states will be incorporated in a more general 
natural computing context with less constraints.
Log-scale encoding of information is shown to play a critical role in bridging the computing models of  
classical and quantum waves. 
Our goal is to redefine AI computing model with quantum-inspired approach and ultimately unify the computing 
model of AI and QC, hoping that this path will bring us to better understand the microscopic computing 
model of the brain.

\section*{Appendix A: Implementation sketch}

\begin{figure}
\centering
\includegraphics[width=12cm, clip]{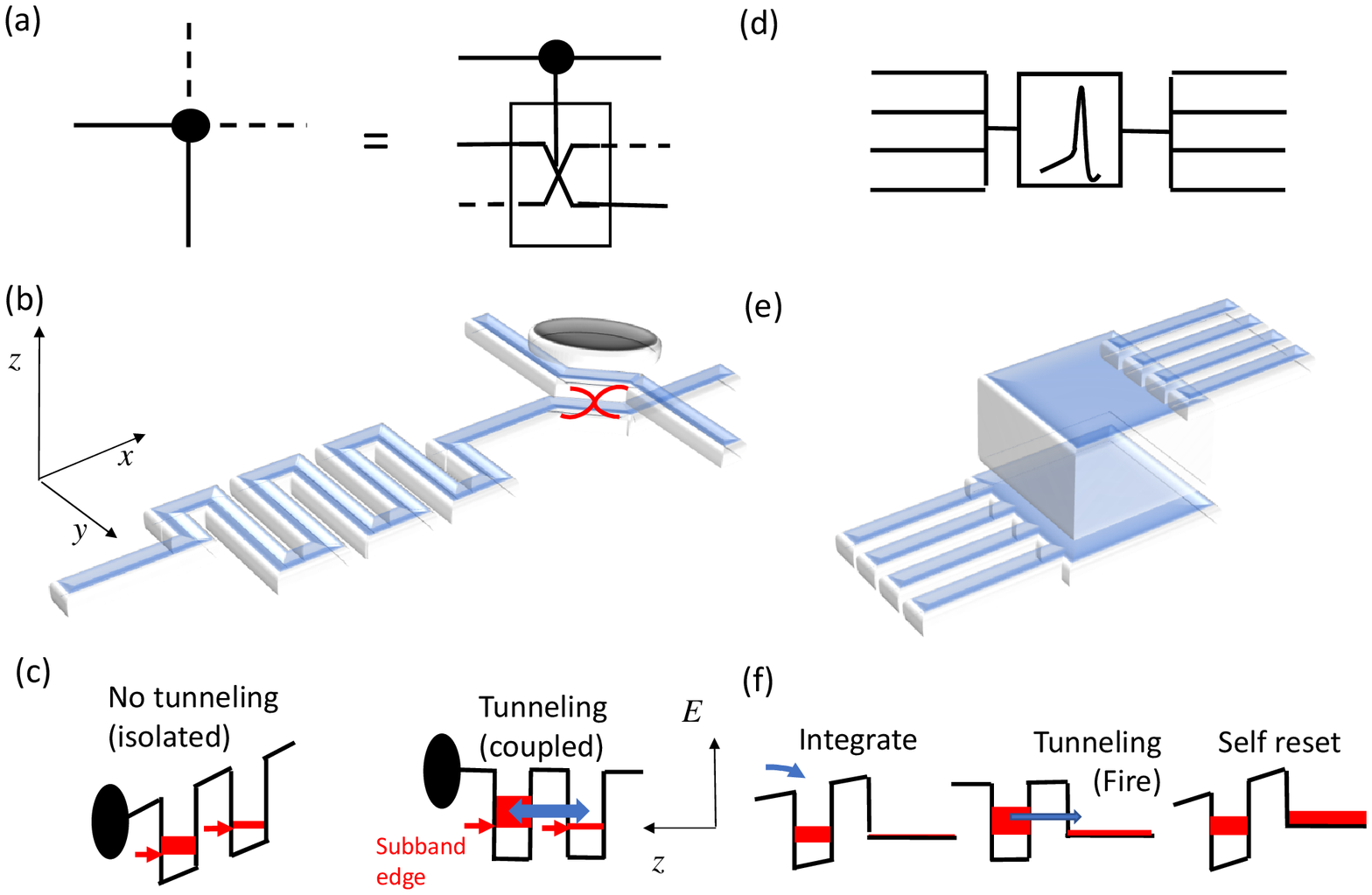}
\caption{Implementation sketch for key primitive building blocks: (a) Weight matrix  
diagram represented by Fredkin gate for cubits; 
(b) Fredkin gate with delay for cubits with nanostructured electron wave guides and utilizes lateral 
tunneling transistor structure \cite{KataAPL};
(c) Energy diagram for lateral tunneling transistor when the subband edges are 
misaligned and decoupled (left) and aligned and coupled (right);
(d) Integrated and fire diagram with combiner and splitter; (e) 
Integrate and fire building block with combiner and splitter with quantum well and wires;
(f) Energy diagram for the integrate and fire building block.}
\label{fig8}
\end{figure}

There are various candidates to implement cubits and their operations, such as 
spin waves, acoustic waves, slow optics, and ion density waves. 
Here, we show implementation sketch for key primitive building blocks using 
nanostructured electronics in Fig. \ref{fig8}. It may have a 
better path to be migrated in existing semiconductor technologies, since 
no signal wave conversion is needed. The purpose is to show that the proposed computing 
model has a reasonable path toward implementation, but further study is needed to identify 
the most suitable way of realizing the device. 
In Fig. \ref{fig8} (a), weight matrix  
diagram is represented by Fredkin gate for cubits.  
It has three inputs: 
\begin{equation}
\ket{\ket{c}} = \bar{c}\ket{\ket{0}}+c\ket{\ket{1}},
\end{equation}
\begin{equation}
\ket{\ket{p}} = \bar{p}\ket{\ket{0}}+p\ket{\ket{1}},
\end{equation}
\begin{equation}
\ket{\ket{c}} = \bar{q}\ket{\ket{0}}+q\ket{\ket{1}}
\end{equation}
for full cubits and 
\begin{equation}
\ket{\ket{c}} = c\ket{\ket{1}},
\end{equation}
\begin{equation}
\ket{\ket{p}} = p\ket{\ket{1}},
\end{equation}
\begin{equation}
\ket{\ket{c}} = q\ket{\ket{1}}
\end{equation}
for half cubits. Note that $\ket{\ket{c}}$ signal may not be spiking. 
The outputs are defined as
\begin{equation}
\ket{\ket{c}} = \bar{c}\ket{\ket{0}}+c\ket{\ket{1}},
\end{equation}
\begin{equation}
\bar{c}\ket{\ket{p}}+c\ket{\ket{q}} = (\bar{c}\bar{p}+c\bar{q})\ket{\ket{0}}+(\bar{c}p+cq)\ket{\ket{1}},
\end{equation}
\begin{equation}
\bar{c}\ket{\ket{q}}+c\ket{\ket{p}} = (\bar{c}\bar{q}+c\bar{p})\ket{\ket{0}}+(\bar{c}q+cp)\ket{\ket{1}}
\end{equation}
for full cubits and 
\begin{equation}
\ket{\ket{c}} = c\ket{\ket{1}},
\end{equation}
\begin{equation}
\bar{c}\ket{\ket{p}}+c\ket{\ket{q}} = (\bar{c}p+cq)\ket{\ket{1}},
\end{equation}
\begin{equation}
\bar{c}\ket{\ket{q}}+c\ket{\ket{p}} = (\bar{c}q+cp)\ket{\ket{1}}
\end{equation}
for half cubits. As indicated in these equations, both superposition (addition) 
and multiplications are involved in the cubit Fredkin gate.
The Fredkin gate is a universal classical gate with constant ancilla inputs \cite{FT}. 
Fredkin gate is more suitable than Toffoli gate for our purpose 
since the number of conserved 0's and 1's between inputs and outputs can lead 
to more efficient passive gate implementations for better energy efficiency. 
Normalized half cubits with $\ket{\ket{0}}$ for no signal and $\ket{\ket{1}}$ for a full signal 
can be assumed. With this assumption, superposed cubit states can be represented by the scalar 
strength of $\ket{\ket{1}}$. Because the Fredkin gate conserves the number of 0's and 1's, 
the gate can be implemented passive wave couplers without active amplification. 
With an active amplification option, the gate can generate signal gain for a larger fanout assuming 
unnormalized cubits. 
When Fredkin gates are combined, the weighted sums of any number of inputs and outputs can 
be flexibly calculated by using analog weighted-sum calculation. 
The data cubits $\ket{\ket{p}}$ and $\ket{\ket{q}}$ for the Fredkin gate 
are repartitioned in accordance with instructions from the control cubit $\ket{\ket{c}}$. When 
the blocks of the gate structure are combined, they can represent any repartitioning 
across a large number of data cubits as a unitary (flux conserving) transformation. 

Figure \ref{fig8} (b) depicts 
Fredkin gate with delay for cubits with nanostructured electron wave guides and utilizes lateral 
tunneling transistor structure \cite{KataAPL, IEDMdemo, KataQW}, 
with energy  diagram for lateral tunneling transistor
in Fig. \ref{fig8} (c).
Here, single-particle electron quantum wave functions 
are treated as Cartesian product classical waves in computing.  
The gate illustrated in Fig. \ref{fig8} is replicated in 
parallel \cite{KataDup} using the lateral tunneling transistor structure. The device consists 
of coupled quantum structures, such as wells and wires, and an alignment gate. 
Without gate bias, the subband edges of the coupled quantum wells and wires are not 
aligned and thus are isolated. Application of an alignment gate voltage adjusts the subband 
edge alignment. When the two subband edges are perfectly aligned, we can make this work 
as 50:50 splitter/combiner by appropriately choosing the design parameters \cite{NANO2019}. 
%since the weights for the wave amplitude 
%can be expressed as 
%\begin{equation}
%cos(\omega_R { L \over v_F}),
%\end{equation}
%and 
%\begin{equation}
%sin(\omega_R { L \over v_F}),
%\end{equation} 
%where $\omega_R$ is the Rabi oscillation frequency, $L$ is the length of the active region defined 
%by the alignment gate, and $v_F$ is the Fermi velocity which corresponds to 
%$v_g$ for electrons. 
The device is driven by quantum physics in $z$ direction \cite{KataQW}
but by classical physics   
in $xy$ direction for ensemble averaging. The structure can be replicated 
for nonbinary weights as expected in Fig. \ref{fig6}. 
%This averaging enables Fredkin gates to perform weighted sum operations cubits for which the ratio 
%is determined by the value of $\ket{\ket{c}}$s. 
The concurrent averaging can be done by using the on-off 
encoding illustrated in Fig. \ref{fig3} (a) such as by using $\ket{0}$ for ``without electrons'' 
and $\ket{1}$ for ``with electrons.'' 
The spatially parallel and temporarily concurrent 
measurement capability for the ensemble averaging is a unique capability of 
cubits in comparison with the measurements for qubits, which have to occur serially with no cloning. 
This feature should be useful in handling large fanout connections as is done by the brain. 
By combining blocks with this structure, a $W_{ij}$ larger than $2 \times 2$ can be constructed. 
Weight update can be performed by changing the gate voltages. The effect of leakage current 
on the on-off ratio is a serious problem for the building blocks in classical computing 
but not in QIC with classical waves. 

Figure \ref{fig8} (d) shows integrated and fire diagram with combiner and splitter. 
Integrate and fire building block with combiner and splitter with quantum well and wires is 
shown in Fig.\ref{fig8} (e) with energy diagram for the integrate and fire building block in 
Fig. \ref{fig8} (f). Details of this block has been presented elsewhere \cite{NANO2019}. 

%\newpage
\section*{Appendix B: Remarks on entanglement}

\begin{figure}
\centering
\includegraphics[width=10cm, clip]{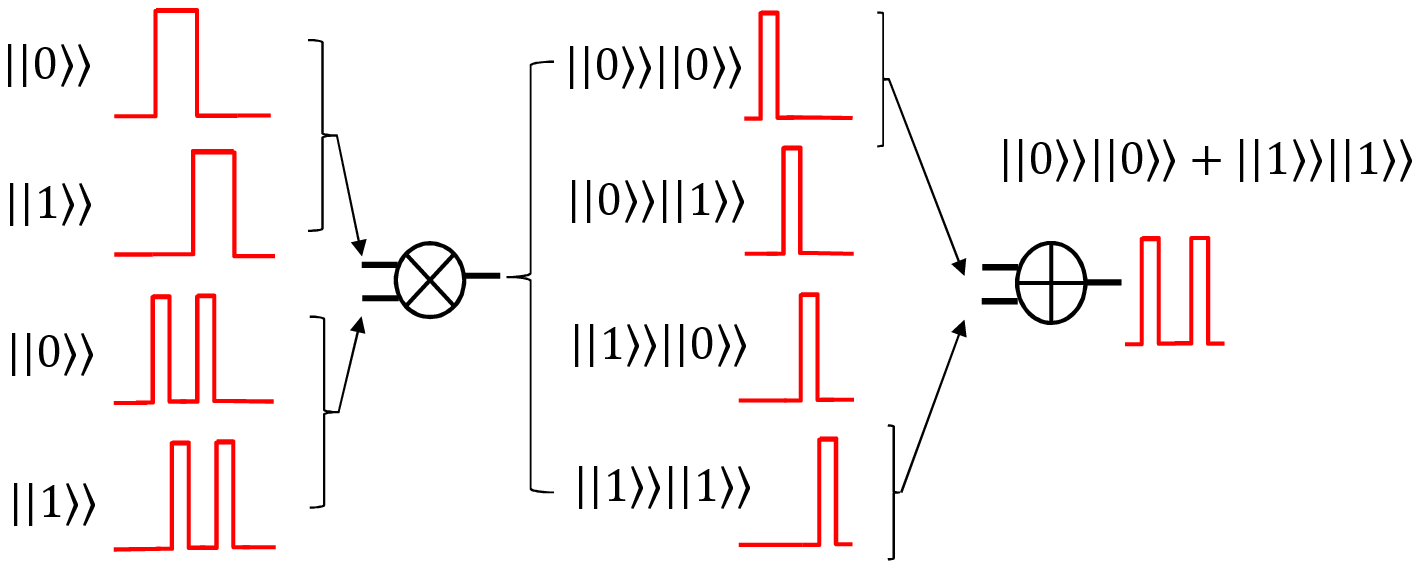}
\caption{Cubit states are encoded into classical waves using the Walsh function. 
Tensor product states and local entanglement can be generated.}
\label{fig9}
\end{figure}

\begin{figure}
\centering
\includegraphics[width=5cm, clip]{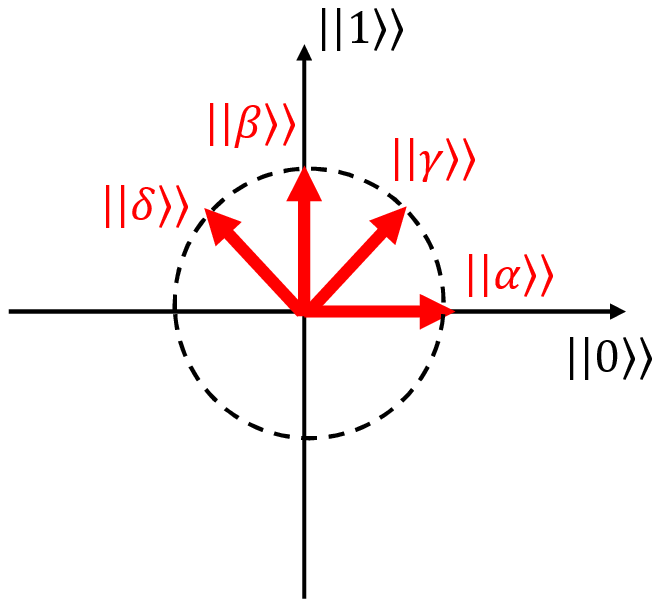}
\caption{Coherent detection of $\ket{\ket{\gamma}}$ and $\ket{\ket{\delta}}$ 
with in-phase $\ket{\ket{\alpha}}$ 
and quadrature $\ket{\ket{\beta}}$. 
%where $I_\gamma = \braket{\braket{\alpha|\gamma}}$ 
%and $Q_\gamma = \braket{\braket{\beta|\gamma}}$.
} 
\label{fig10}
\end{figure}

Finally, let us spend some time to discuss 
how much QIC can be closer to QC. 
Multiple cubit states can be represented in tensor products by using multipliers. 
For example, the Walsh function can be used to represent such tensor product states as those 
illustrated in Fig. \ref{fig9}. For simplicity, it is assumed that the physical pulses 
have square shapes in an unlimited bandwidth environment. 
Entangled states such as 
\begin{equation}
{1 \over \sqrt{2}}(\ket{\ket{0}}\ket{\ket{0}}+\ket{\ket{1}}\ket{\ket{1}})
\end{equation} 
can be generated as a linear combination of the tensor product states. 
Since this is a local entanglement, it differs from the nonlocal entanglement in qubits. 
Tensor product states can be constructed in spatial degrees of freedom as well. Since the brain 
occupies relatively small amount of the space, this local entanglement feature can 
provide a similar feature with nonlocal quantum entanglement \cite{BQC6}. 
Fig. \ref{fig10} shows four states, $\ket{\ket{\alpha}} = \ket{\ket{0}}$, 
$\ket{\ket{\beta}} = \ket{\ket{1}}$, $\ket{\ket{\gamma}} = 1/\sqrt{2} (\ket{\ket{0}} 
+ \ket{\ket{1}})$, and $\ket{\delta} = 1/\sqrt{2} (-\ket{\ket{0}} + \ket{\ket{1}})$, created by using 
the phase encoding in Fig. \ref{fig2} (b), presumably 
with a carrier. Given these four states, 
\begin{equation}
D(\ket{\ket{\alpha}}, \ket{\ket{\gamma}}) - D(\ket{\ket{\alpha}}, \ket{\ket{\delta}}) +
D(\ket{\ket{\beta}}, \ket{\ket{\gamma}}) + D(\ket{\ket{\beta}}, \ket{\ket{\delta}}) 
= 2\sqrt{2},
\end{equation}
when $D(\ket{\ket{x}}, \ket{\ket{y}}) = 
\braket{\braket{x|y}}$ as is commonly used 
in wireless coherent detection algorithms \cite{Wireless}. 
This argument may suggest that this relation is arising from wave nature, not necessarily quantum nature.

%\newpage
\begin{addendum}
\item The author is grateful to colleagues inside and outside IBM Research for their valuable discussions. 
\item[Competing Interests] The author declares no competing interests.
\item[Correspondence] Correspondence and requests for materials should be addressed to Yasunao Katayama~(email: yasunaok@jp.ibm.com).
\end{addendum}

\end{document}